\documentclass[10pt, pre, eps, twocolumn,superscriptaddress,showpacs,superscriptaddress]{revtex4-1}
\usepackage{amsmath,amssymb,amsbsy,color,graphicx}
\usepackage{fourier}
\usepackage{ulem}

\newcommand{\ave}[1]{\langle #1 \rangle}

\newcommand{\dfracp}[2]{\dfrac{\partial #1}{\partial #2}}
\newcommand{\II}[1]{\ave{\ave{#1}}}

\begin{document}

\title{Non-mean-field critical exponent in a mean-field model :
Dynamics versus statistical mechanics}

\author{Shun Ogawa}

\affiliation{Department of Applied Mathematics and Physics,
Graduate School of Informatics, Kyoto University, Kyoto, 606-8501, Japan}

\author{Aurelio Patelli}
\affiliation{Dipartimento di Fisica e Astronomia, Universit\'a di Firenze and INFN, Via Sansone 1, IT-50019 Sesto Fiorentino, Italy}

\author{Yoshiyuki Y. Yamaguchi}
\affiliation{Department of Applied Mathematics and Physics,
Graduate School of Informatics, Kyoto University, Kyoto, 606-8501, Japan}

\date{\today}
\begin{abstract}
 	The mean-field theory tells that the classical critical exponent
        of susceptibility is the twice of that of magnetization.
	However, the linear response theory
      	based on the Vlasov equation,
      	which is naturally introduced by the mean-field nature,
      	makes the former exponent half of the latter
      	for families of quasistationary states
      	having second order phase transitions
      	in the Hamiltonian mean-field model
      	and its variances. 
    	We clarify that this strange exponent is due to existence
    	of Casimir invariants which trap the system in a quasistationary state
    	for a time scale diverging with the system size.
    	The theoretical prediction is numerically confirmed
    	by $N$-body simulations for the equilibrium states
        and a family of quasistationary states.
\end{abstract}

\maketitle

\section{Introduction}
\label{sec:introduction}
Are critical exponents of an isolated dynamical system 
the same as ones computed via the statistical mechanics?
We tackle this question, by dealing with a ferromagnetic like model
in the mean-field universality class and considering the critical 
exponents of the zero-field susceptibility. 
Isothermal susceptibility, $\chi^{\rm T}$,
can be obtained by using standard methods of statistical mechanics,
while the susceptibility of an isolated system, $\chi^{\rm I}$,
can be derived from linear response theory \cite{Kubo}.
These two susceptibilities satisfy the inequality
$\chi^{\rm I}\leq\chi^{\rm T}$ \cite{chis1,chis2}
which is derived considering existence of invariants
\cite{mazur-69,suzuki-70}.
This implies that the exponents, $\gamma^{\rm T}$ and $\gamma^{\rm I}$,
with which the two susceptibilities diverge at the critical point,
satisfy $\gamma^{\rm I}\leq\gamma^{\rm T}$.
Is it possible that $\gamma^{\rm I}$ is strictly smaller than $\gamma^{\rm T}$?
A difficulty in answering this question is
that the susceptibility of an isolated system cannot be easily evaluated.
In this article we show how kinetic theory \cite{balescu,nicholson}
can effectively answer the initial question
in systems of the mean-field type
using a recently developed version of linear response theory
\cite{LinearResponseI,LinearResponseII}
based on the Vlasov equation. 

Many different physical systems can be described by kinetic theory,
including self-gravitating systems, plasmas and fluids \cite{balescu,binney,nicholson}.
For $N$-particle systems with long-range interactions \cite{campa-dauxois-ruffo-09}
both perturbative approaches \cite{balescu}
and the rigorous mean-field limit \cite{spohn-book,braun}
lead to a description of the system in the continuum, $N\to\infty$,
limit in terms of the Vlasov equation. 
This equation rules the time-evolution
of the single-particle distribution function
and has an infinity of stationary solutions.  
For instance, all distribution functions
that depend on phase-space variables
only through the single-particle energy do not evolve in time,
as proven by Jeans \cite{Jeans-1919}.
On the long time scale, the system is described
by appropriate kinetic equations which include ``collisional'' (finite $N$)
effects, like Landau and Balescu-Lenard equation,
and evolves towards Boltzmann-Gibbs (BG) equilibrium. 
However, since the relaxation time scale diverges with $N$ \cite{physicaa}, 
the early evolution of the system is well described by the Vlasov equation.
Therefore, the use of the linear response theory developed in \cite{LinearResponseI,LinearResponseII}
is appropriate in the large $N$ limit.

In order to perform explicit calculations of susceptibilities,
it is convenient to consider the so-called Hamiltonian Mean-Field (HMF) model
\cite{campa-dauxois-ruffo-09,spohn,inagaki-konishi-93,antoni-ruffo-95}.
This model describes the motion of $N$ particles on a circle interacting
with an attractive cosine potential.
The BG equilibrium solution of this model displays
a high-energy phase where the particles are uniformly distributed
on the circle and a low-energy phase where the particles form a cluster. 
The two phases are separated by a second order phase transition point
at which susceptibility diverges with the classical mean-field exponents.
On the other hand, in the mean-field limit,
the time-evolution of the single-particle distribution function
of the HMF model is exactly described by the Vlasov equation. 
Moreover, a BG homogeneous state is a stationary solution of this equation
which looses its stability at an energy which coincides with the 
second-order phase transition energy \cite{inagaki-konishi-93}. 
Below this energy, the BG inhomogeneous state is
also a stable stationary state of the Vlasov equation. 

Stable stationary states almost do not evolve
even in the system with finite but large $N$,
and called as quasistationary states \cite{physicaa,physicaaII}.
The long-lasting quasistationary states, therefore,
show nonequilibrium phase transitions,
and a phase diagram is theoretically drawn for a set of initial states
with the aid of a nonequilibrium statistical mechanics \cite{antoniazzi-07}.
In this article, we perform a detailed analysis of the scaling laws
of susceptibility around the critical point
of (non)equilibrium phase transitions for quasistationary states.
We remark that the BG equilibrium states are ones of quasistationary states,
and hence the obtained scaling laws are valid even for the BG equilibrium states.

This article is constructed as follows.
We introduce the HMF model and the corresponding Vlasov system
in Sec.~\ref{sec:HMFmodel}.
The scaling of the Vlasov susceptibility is analyzed in Sec.~\ref{sec:scaling},
and the theoretical prediction is numerically confirmed
in Sec.~\ref{sec:numerics}.
A generalization from the HMF model is discussed
in Sec.~\ref{sec:generalization}.
Sec.~\ref{sec:summary} is devoted to summary and discussions.

\section{Hamiltonian mean-field model}
\label{sec:HMFmodel}
The Hamiltonian function of the HMF model reads
\begin{equation}
    \begin{split}
        H_{N}
        & = \sum_{i=1}^{N} \left[
          \dfrac{p_{i}^{2}}{2} 
          + \sum_{j=1}^{N} \dfrac{1-\cos(q_{i}-q_{j})}{2N} 
          - h\Theta(t) \cos (q_{i}-\phi) \right]
    \end{split}
\end{equation}
where $h$ and $\phi$ are respectively the modulus and the phase
of the external magnetic vector $(h\cos\phi,h\sin\phi)$,
and $\Theta(t)$ is the Heaviside step function.
The magnetization vector $(\ave{M_{x}}_{N},\ave{M_{y}}_{N})$ is defined by
\begin{equation}
    (\ave{M_{x}}_{N},\ave{M_{y}}_{N})
    = \dfrac{1}{N} \sum_{j=1}^{N} (\cos q_{j},\sin q_{j}),
\end{equation}
where $\ave{\cdot}_{N}$ represents the average over $N$ particles.
The isolated system ($h=0$) has the rotational symmetry,
therefore, we consider $\phi=0$ and $\ave{M_{y}}_{N}=0$
without loss of generality. As a consequence, 
we call the $x$-axis the direction of the spontaneous magnetization.

The corresponding effective one-particle Hamiltonian of HMF is
\begin{equation}
    \label{eq:Hf}
    H_{h}[f_{h}](q,p,t) = \dfrac{p^{2}}{2} - \ave{M}_{h}\cos q
    - h\Theta(t)\cos q,
\end{equation}
where the magnetization observable is $M(q)=\cos q$
and the brackets $\langle\cdots\rangle_h$ means
the average respects to the single particle distribution $f_h$.
The distribution $f_{h}$ evolves 
following the Vlasov equation
\begin{equation}
    \dfracp{f_{h}}{t} + \{ H_{h}[f_{h}], f_{h} \} = 0,
	\label{eq:vlasov}
\end{equation}
where $\{\cdot,\cdot\}$ is the Poisson bracket defined by
\begin{equation}
    \{a,b\} = \dfracp{a}{p} \dfracp{b}{q} - \dfracp{a}{q} \dfracp{b}{q}.
\end{equation}
We note that the magnetization $\ave{M}_{h}$ appearing in $H_{h}$,
\eqref{eq:Hf}, is determined self-consistently to satisfy the equation
\begin{equation}
    \label{eq:self-consistent}
    \ave{M}_{h}= \iint \cos q f_{h} dqdp.
\end{equation}

Let us consider the case in which the external field is turned off.
In that case all the stationary states are in the Jeans' class,
and are functions of $H_{0}$ specified by
\begin{equation}
    \label{eq:f0}
    f_{0}(q,p;b,a_{1},\cdots,a_{n})
    = \dfrac{F(H_{0}(q,p);b,a_{1},\cdots,a_{n})}
    {\II{F(H_{0}(q,p);b,a_{1},\cdots,a_{n})}},
\end{equation}
where
\begin{equation}
    \II{\varphi(q,p)}=\iint \varphi(q,p) dqdp.
\end{equation}
For instance, in the canonical equilibrium, 
the function $F$ of energy is 
$F(E;b)=e^{-bE}$, where $b$ is the inverse temperature.
For the sake of simplicity,
we consider only one independent parameter $b$,
and other parameters $a_{1},\cdots,a_{n}$ depend on $b$.
The effective Hamiltonian of any one-dimensional system in a stationary state is integrable, 
and the angle-action variables $(\theta,J)$ \cite{BOY10}
can be introduced accordingly. 
The Hamiltonian $H_{0}[f_{0}]$ and the distribution function $f_{0}$ 
depend only on the action $J$.

\section{Scaling of the Vlasov susceptibility}
\label{sec:scaling}

The Vlasov susceptibility is given by the linear response theory
\cite{LinearResponseI,LinearResponseII},
and reads
\begin{equation}
    \label{eq:chi-Vlasov}
    \chi^{\rm V}(b) = \dfrac{1-D^{\rm V}(b)}{D^{\rm V}(b)},
\end{equation}
where $D^{\rm V}$ is the stability functional,
and $D^{\rm V}(b)>0$ implies the stability of the state
\cite{physicaa,ogawa13}.
This functional can be decomposed in two terms
\begin{equation}
    \label{eq:D-Vlasov}
    D^{\rm V}(b) = D^{\rm V}_{1}(b) + D^{\rm V}_{2}(b),
\end{equation}
where the first one is
\begin{equation}
    D^{\rm V}_{1}(b) = 1 + \dfrac{\II{F'(H_{0}(J);b)\ave{\cos^{2}q}_{J}}}
    {\II{F(H_{0}(J);b)}},
\end{equation}
while the second one is
\begin{equation}
    D^{\rm V}_{2}(b) = - \dfrac{\II{F'(H_{0}(J);b)\ave{\cos q}_{J}^{2}}}
    {\II{F(H_{0}(J);b)}}.
\end{equation}
The prime means the derivative $F'=dF/dE$ and
$\ave{\cdots}_{J}$ represents the average 
with fixed $J$, i.e.,
\begin{equation}
    \label{eq:avetheta}
    \ave{\varphi(\theta,J)}_{J}
    = \dfrac{1}{2\pi} \int_{-\pi}^{\pi} \varphi(\theta,J)~ d\theta.
\end{equation}
For homogeneous distribution, we have $q=\theta$ and $D^{\rm V}_{2}$ vanishes.
The stability functional in this case is
\begin{equation}
    \label{eq:D-Vlasov-homo}
    D^{\rm V}_{\rm homo}(b) = 1 + \dfrac{1}{2}
    \dfrac{\II{F'(p^{2}/2;b)}}{\II{F(p^{2}/2;b)}}.
\end{equation}
For instance, using the canonical equilibrium, we obtain that
the susceptibility is $\chi^{\rm V}=b/(b-2)$
and its critical point is $b_{\rm c}^{\rm cano}=2$.

Let us introduce three assumptions in order to
obtain Jeans' distributions \eqref{eq:f0} which describe 
continuous phase 
transitions at $b=b_{\rm c}$: 
(I) The states are homogeneously stable for $b<b_{\rm c}$
and inhomogeneously stable for $b>b_{\rm c}$.
(II) The magnetization $\ave{M}_{0}$ is a continuous function of $b$.
(III) The solution of the self-consistency equation \eqref{eq:self-consistent} 
gives an unstable homogeneous branch for $b>b_{\rm c}$.
As a consequence, the stability functional \eqref{eq:D-Vlasov}
is positive for any $b (\neq b_{\rm c})$, and
\begin{equation}
    \ave{M}_{0} = \left\{
      \begin{array}{ll}
          0 & \quad b\leq b_{\rm c}, \\
          (b-b_{\rm c})^{\beta} &\quad b\agt b_{\rm c}.
      \end{array}
    \right.
\end{equation}
Moreover, in the homogeneous branches of both sides,
the stability functional reads
\begin{equation}
    \label{eq:Dhomo-cond-1}
    D^{\rm V}_{\rm homo}(b) = \left\{
      \begin{array}{ll}
          c_{+}(b)(b_{\rm c}-b)^{\Gamma_{+}}, & b\alt b_{\rm c}, \\
          -c_{-}(b)(b-b_{\rm c})^{\Gamma_{-}}, & b\agt b_{\rm c}
      \end{array}
    \right.
\end{equation}
with positive $c_{\pm}(b)$ and $\pm$ discriminates
between the two regions over-critical and under-critical.
The exponents $\Gamma_{\pm}$ depend on the choice 
of the parameter $b$ of the distribution.
In general we can consider a parametrization such that $\Gamma_{\pm}=1$.

The critical exponents, $\gamma^{\rm V}_{\pm}$,
of the susceptibility \eqref{eq:chi-Vlasov} depend
on the behavior of the stability functional
$D^{\rm V}(b)\to 0$ close to the critical point $b_{\rm c}$.
Equation \eqref{eq:Dhomo-cond-1} gives
$\gamma^{\rm V}_{+}=\Gamma_{+}$ 
when the state of the system is homogeneous,
that is equal to the classical exponent.
In the following, we show the non-classical relation
$\gamma^{\rm V}_{-}=\beta/2$ settled in the inhomogeneous phase,
where $\beta=\Gamma_{-}/2$ in the case ferromagnetic mean-field systems.

Let us start showing the relation $\beta=\Gamma_{-}/2$.
Around the critical point $b\gtrsim b_{\rm c}$,
the magnetization $\ave{M}_{0}$ is small by assumption (II),
and we expand $F(H_{0})$ around the critical point as
\begin{equation}
    \label{eq:F-expand}
    F(H_{0};b)
    = \sum_{n=0}^{\infty} \dfrac{(-\ave{M}_{0}\cos q)^{n}}{n!} F^{(n)}(p^{2}/2;b),
\end{equation}
where $F^{(n)}$ is the $n$-th derivative of $F$
and we assumed that $f_{0}$ depends on $\ave{M}_{0}$
through $H_{0}$ only. For such
distributions,
the self-consistency equation \eqref{eq:self-consistent} becomes
\begin{equation}
    A_{1}(b)\ave{M}_{0} + A_{3}(b)\ave{M}_{0}^{3} + O(\ave{M}_{0}^{5}) = 0,
\end{equation}
where
\begin{equation}
    \begin{split}
        & A_{1}(b) = 1 - \dfrac{B_{1}}{B_{0}} = D^{\rm V}_{\rm homo}(b), \\
        & A_{3}(b) = \dfrac{B_{1}B_{2}-B_{0}B_{3}}{B_{0}^{2}},
    \end{split}
\end{equation}
and $B_{n}~(n=0,1,2,\cdots)$ are defined by
\begin{equation}
    B_{n}(b) = \dfrac{(-1)^{n}}{n!} \iint F^{(n)}(p^{2}/2;b) \cos^{2\lceil n/2 \rceil}q dqdp,
\end{equation}
with $\lceil x \rceil=\min\{m\in\mathbb{Z}|m\geq x\}$.
We further assume that $B_{n}(b_{\rm c})\neq 0$ for any $n$.
The non-zero solution of the self-consistency equation gives the scaling
\begin{equation}
    \ave{M}_{0} = \sqrt{\dfrac{c_{-}(b)}{A_{3}(b)}}~ (b-b_{\rm c})^{\Gamma_{-}/2}
\end{equation}
whenever $A_{3}(b)>0$,
which implies existence of Jeans' inhomogeneous states.
We, therefore, get the relation $\beta=\Gamma_{-}/2$.

To prove the main relation $\gamma^{\rm V}_{-}=\beta/2$,
we separately estimate $D^{\rm V}_{1}$ and $D^{\rm V}_{2}$.
To evaluate the behavior of the first term
we remark that $\II{F'(H_{0};b)\ave{\cos^{2}q}_{J}}=\II{F'(H_{0};b)\cos^{2}q}$.
Using the expansion \eqref{eq:F-expand},
the first component of the stability functional
scales as $D^{\rm V}_{1}(b)\sim (b-b_{\rm c})^{\Gamma_{-}}$ for $b>b_{\rm c}$.
The second component is given by \cite{LinearResponseII}
\begin{equation}
    D^{\rm V}_{2}(b) = 16\sqrt{\ave{M}_{0}}~(I_{1}+I_{2}),
\end{equation}
where
\begin{equation}
    I_{1} = - \int_{0}^{1}
    \left[ \dfrac{2E(k)}{K(k)} - 1 \right]^{2}
    kK(k)\frac{F'(\ave{M}_{0}(2k^{2}-1))}{\II{F(H_0)}} dk,
\end{equation}
\begin{equation}
    I_{2} = - \int_{1}^{\infty}
    \left[ \dfrac{2k^{2}E(1/k)}{K(1/k)} - 2k^{2}+1 \right]^{2}
    K(1/k) \frac{F'(\ave{M}_{0}(2k^{2}-1))}{\II{F(H_0)}}dk,
\end{equation}
and $K$ and $E$ are respectively the complete elliptic integrals
of the 1st and the 2nd kinds.
The integrals $I_{1}$ and $I_{2}$ converge to non-zero constants
in general even in the limit $\ave{M}_{0}\to 0$. 
Hence, the second part scales as
\begin{equation}
    D^{\rm V}_{2}(b)\sim\sqrt{\ave{M}_{0}}\sim (b-b_{\rm c})^{\beta/2}.
\end{equation}
Close to the critical point the second component $D^{\rm V}_{2}$ dominates 
since it goes to zero slower compared with the first one $D^{\rm V}_{1}$.
Consequently, the Vlasov critical exponents for Jeans' distributions are
\begin{equation}
    \label{eq:scaling-relation}
    \gamma^{\rm V}_{-}=\beta/2=1/4, \quad
    \gamma^{\rm V}_{+}=1,
\end{equation}
when $\Gamma_\pm=1$.
We stress that the exponent $\gamma^{\rm V}_{-}=\beta/2=1/4$ differs
from the classical $\gamma_{-}=1$ \cite{chavanis-11}.

We explain that the {\it strange} exponent $\gamma^{\rm V}_{-}=\beta/2$
is due to infinite invariants of the Vlasov equation,
called Casimirs.
A Casimir is a functional of the distribution function $\int s(f)dqdp$,
where $s$ is any smooth function.
It is an integral of motions of the Vlasov dynamics 
whenever the distribution solves the Vlasov equation \eqref{eq:vlasov} itself.
As a consequence, any Casimir introduce a conservation law
and the second component of the stability functional, 
which gives the {\it strange} exponent, 
takes care of whole of them.

The variation of the distribution $\delta f=f_{h}-f_{0}$ satisfies
\begin{equation}
    0 = \iint [s(f_{0}+\delta f)-s(f_{0})] dqdp
    = \int s'(f_{0}(J)) \widetilde{\delta f}_{0}(J) dJ
\end{equation}
up to the linear order,
where $\widetilde{\delta f}_{0}(J)$ is the Fourier zero mode of $\delta f$
with respect to the angle $\theta$. This constraint must hold for any smooth
functions $s$, and hence $\widetilde{\delta f}_{0}(J)=0$ \cite{ogawa13}.
Let us derive $f_{h}$
from the test function with the external field,
\begin{equation}
    \label{eq:gh}
      g_{h}(q,p;b) = \dfrac{F_{h}(H_{h}(q,p;b))}{\II{F_{h}(H_{h}(q,p;b))}},
\end{equation}
where $F_{h}$ is a family of functions of energy and is expanded as
\begin{equation}
    F_{h} = F + h G + O(h^{2}).
\end{equation}
By the definition of the susceptibility $\chi^{\rm V}$,
the magnetization $\ave{M}_{h}$ is written as
$\ave{M}_{h} = \ave{M}_{0} + h \chi^{\rm V} + O(h^{2})$,
and hence
\begin{equation}
    H_{h} = H_{0} + h\psi(q) + O(h^{2}),
\end{equation}
where
\begin{equation}
    \psi(q) = - (\chi^{\rm V}+1)\cos q + O(h^{2}).
\end{equation}
Substituting the above two expansions into $g_{h}$, 
and ignoring the term of order $O(h^{2})$, we have
\begin{equation}
    g_{h} = f_{0}
    + h \left[ \dfrac{F'(H_{0})\psi + G(H_{0})}{\II{F(H_{0})}}
      - \dfrac{\II{F'(H_{0})\psi+G(H_{0})} F(H_{0})}{\II{F(H_{0})}^{2}}
    \right].
\end{equation}
Subtracting the Fourier zero mode from $g_{h}-f_{0}$,
the variation must satisfy
$\delta f= g_{h}-f_{0} - \ave{g_{h}-f_{0}}_{J}=g_{h}-\ave{g_{h}}_{J}$ 
and hence
\begin{equation}
    f_{h} = f_{0} - \dfrac{h(\chi^{\rm V}+1)}{\II{F(H_{0})}} F'(H_{0})
    \left( \cos q - \ave{\cos q}_{J} \right),
\end{equation}
where we used $\ave{F(H_{0})}_{J}=F(H_{0})$ and $\ave{G(H_{0})}_{J}=G(H_{0})$
since $H_{0}$ depends on the action $J$ only.
Multiplying by $M(q)=\cos q$ and integrating in the $\mu$ space, 
we get the Vlasov susceptibility \eqref{eq:chi-Vlasov}
and the stability functional \eqref{eq:D-Vlasov}.

Following these results,
we propose a scenario of relaxation as follows
\cite{physicaa,physicaaII}:
When the external field is switched on, the system gets trapped in a QSS
to keep Casimirs invariants 
and this trapping gives the {\it strange} critical exponent
$\gamma_{-}^{\rm V}=\beta/2$.
However, the Vlasov dynamics is not the true dynamics for finite systems,
thus Casimirs are not exactly conserved
but evolve on a time-scale which diverges with $N$.
Consequently, the system goes to the Boltzmann-Gibbs equilibrium
recovering the classical exponent after the equilibration.

\section{Numerical tests}
\label{sec:numerics}

\begin{figure}[t]
    \centering
    \includegraphics[width=9cm]{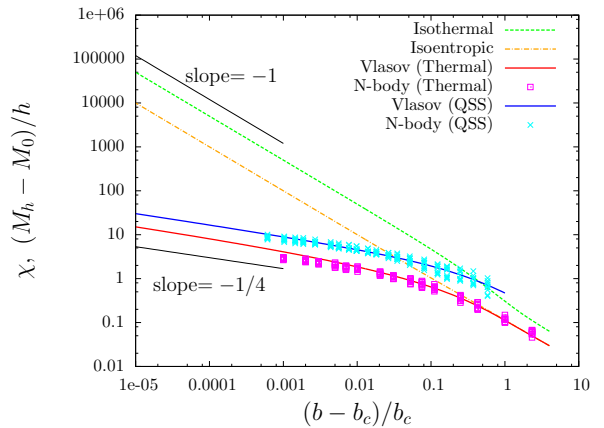}
    \caption{(color online)
        Susceptibilities as functions of the normalized parameter
        $(b_{\rm c}-b)/b_{\rm c}$ in log-log plot.
        Lines report theoretical predictions of
        the isothermal $\chi^{\rm T}$ (green broken),
        the isoentropic $\chi^{\rm S}$ (orange dashed)
        and the Vlasov $\chi^{\rm V}$ (red lower solid)
        susceptibilities for the thermal equilibrium family.
        We remark that $\chi^{\rm S}$ is computed explicitly
        by using the exact solution in the microcanonical statistics
        \cite{campa-dauxois-ruffo-09}
        and by taking the invariance of the entropy during
        the quasistatic adiabatic process into account.
        The Vlasov susceptibility for a QSS family
        of the Fermi-Dirac type is also reported (blue upper solid).
        Points are computed in $N$-body simulations
        and represent $(M_{h}-M_{0})/h$,
        where $M_{h}$ is time average in the period of $t\in [0,500]$.
        $N=10^{6}$ and $h=10^{-2}$ for the thermal equilibrium family
        (purple square),
        and $N=10^{7}$ and $h=10^{-3}$ for the QSS family 
        (light blue cross).
        For each $b$, $10$ points are plotted
        corresponding to $10$ realizations.
    }
    \label{fig:exponent}
\end{figure}

The Vlasov exponent is verified by $N$-body simulations,
which are performed by the 4th order symplectic integrator \cite{yoshida-93}
with the time step $\Delta t=0.1$.
We compute susceptibility for two families of Jeans' class states.
One is the thermal equilibrium
\begin{equation}
    F(E) = e^{-E/T},
\end{equation}
whose control parameter is $b=1/T$ and the critical point is $T_{\rm c}=1/2$.
The other is Fermi-Dirac type
\begin{equation}
    F(E)=\dfrac{1}{e^{(E-\mu)/T}+1}
\end{equation}
with fixed $T=1/5$, whose control parameter is $b=-\mu$
and the critical point is $\mu_{\rm c}\simeq 0.239346$.
The latter is an example of a family of 
out of equilibrium quasistationary states (QSSs),
and the critical exponent $\beta$ is confirmed as $1/2$
by solving the self-consistent equation \eqref{eq:self-consistent}.
The Fermi-Dirac type families are obtained, approximately at least,
by starting from waterbag initial states.
The values of parameters $\mu$ and $T$ are controlled
by suitably choosing the waterbag initial states
with the aid of a nonequilibrium statistical mechanics
\cite{antoniazzi-07}.
For both cases, the Vlasov predictions are in good agreements with 
the $N$-body simulations for time-scales shorter than 
the equilibration one, as shown in Fig.\ref{fig:exponent}.

\begin{figure}
    \centering
    \includegraphics[width=9cm]{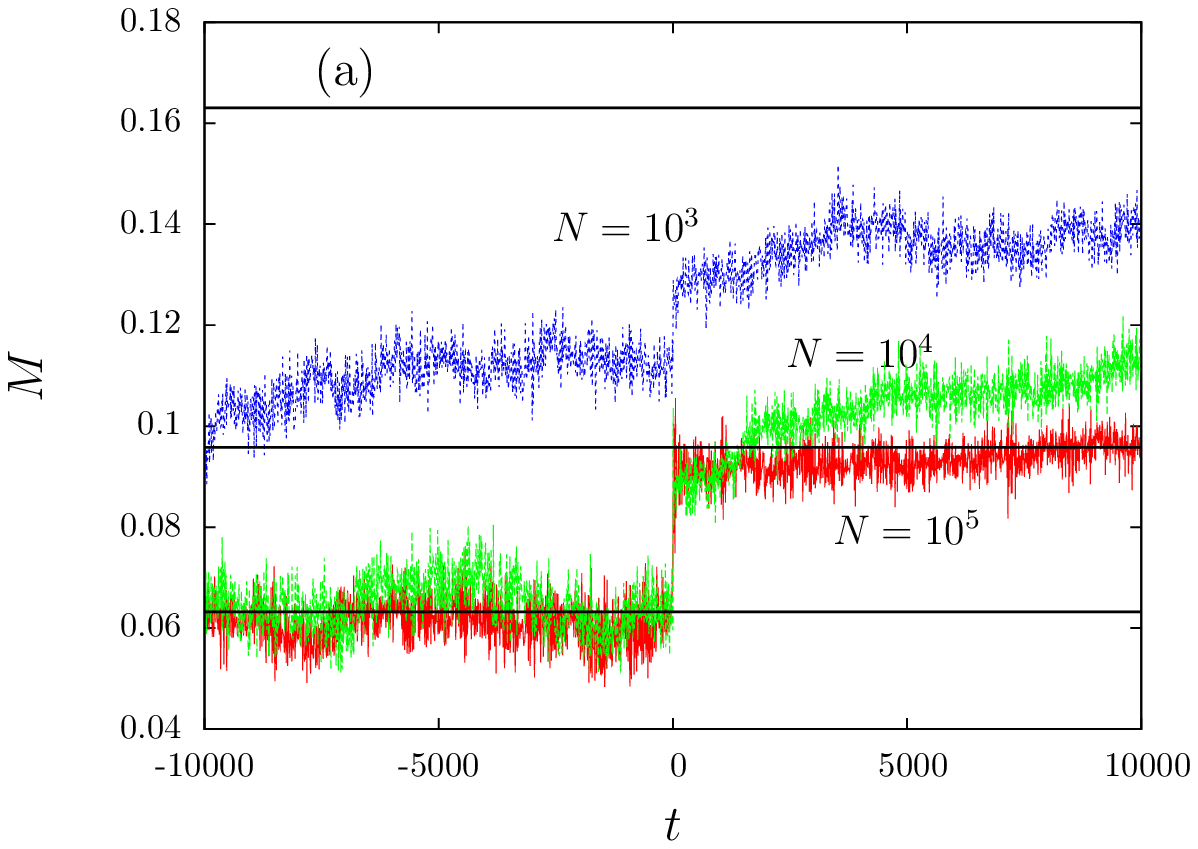}
    \includegraphics[width=9cm]{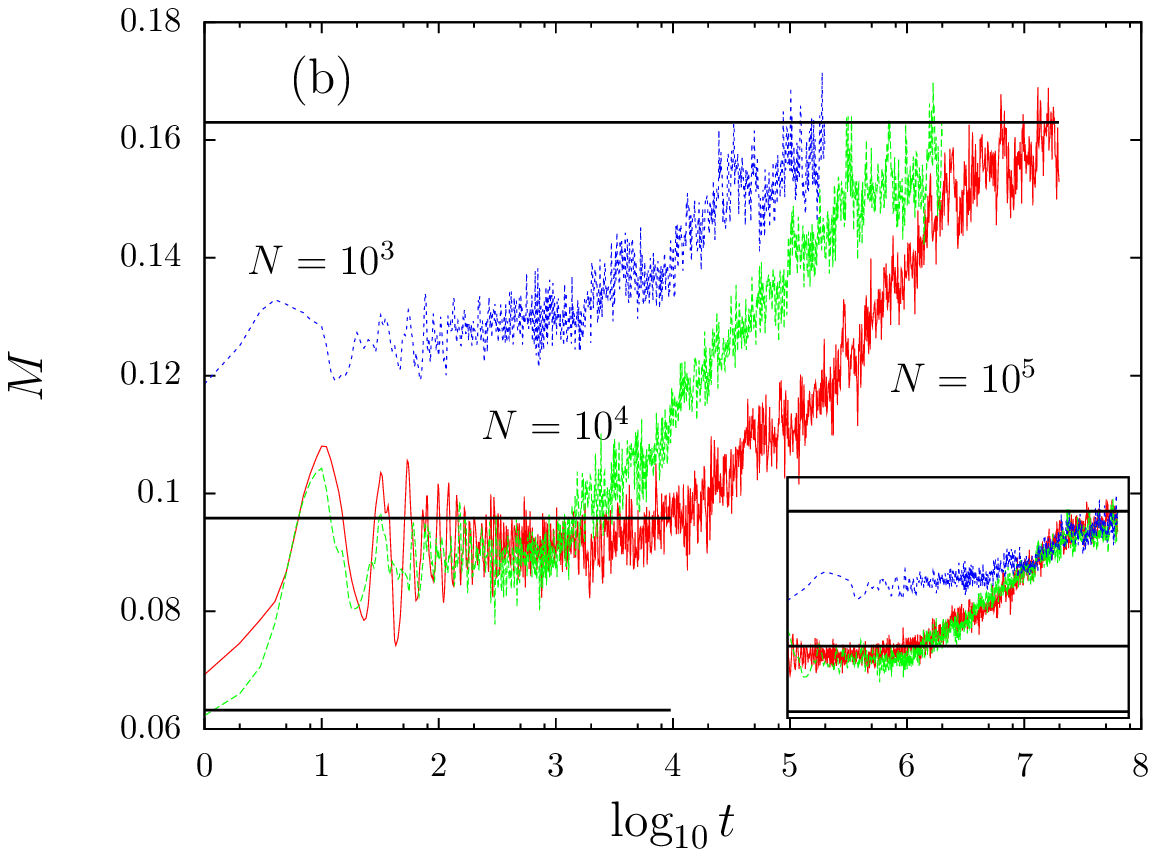}
    \caption{(color online)
      $N$-body simulations in the HMF model with external field.
      (a) Short time evolutions of magnetization.
      (b) Long time evolutions.
      The horizontal axis is in the logarithmic scale,
      which is scaled as $\log_{10}(t/N)$ in the inset. 
      $N=10^{3}~(100)$, $10^{4}~(10)$ and $10^{5}~(1)$,
      where the inside of braces is the number of realizations
      over which the orbits are averaged.
      The system is in thermal equilibrium with $T=0.499$ in $t<0$,
      and the external field turns on with $h=0.01$ at $t=0$.
      The direction of magnetization vector $(M_{x},M_{y})$
      is reset to $x$-direction at $t=0$.
      In each panel, three horizontal lines represent
      equilibrium level with the quasistatic adiabatic susceptibility (upper),
      the QSS level predicted by the Vlasov linear response theory (middle)
      and thermal equilibrium level without external field (lower).}
    \label{fig:Nbody}
\end{figure}

The scenario of relaxation proposed in the last of Sec.\ref{sec:scaling}
is examined by direct $N$-body simulations, shown
in Fig. \ref{fig:Nbody}.
For $t<0$, the system is at equilibrium with a temperature
$T=0.499<1/2=T_{\rm c}$.
The external field with a small magnitude $h=0.01$ is switched on at $t=0$,
and the system jumps to the QSS predicted by 
the linear response theory based on the Vlasov equation.
In the long time regime Casimirs are no more invariants
due to the presence of rare collisions \cite{balescu},
and the system goes towards equilibrium.
Simulations indicate that the time-scale of relaxation from the QSS
to equilibrium grows linearly with $N$,
as found for isolated inhomogeneous QSSs in Ref. \cite{debuyl-mukamel-ruffo-11}.

\section{Generalization of systems}
\label{sec:generalization}

For simplicity, we have concentrated in the HMF model,
but the present theory can be applied to generalized systems.
Let us consider the Hamiltonian
\begin{equation}
    \label{eq:HMFgeneral}
    \begin{split}
        H_{N}
        & = \sum_{i=1}^{N} \dfrac{p_{i}^{2}}{2}
        +\dfrac{1}{2} \sum_{i,j=1}^{N} K_{N}(r_{i}-r_{j})\left( 1- \cos (q_{i}-q_{j})\right) \\
        & - \sum_{i=1}^{N} h_{r_{i}}(t) \Theta(t) \cos q_{i},
    \end{split}
\end{equation}
where $r_{i}$ is the $i$-th lattice point on the one-dimensional lattice,
$r_{i+1}-r_{i}=1$,
the lattice has the periodic boundary condition
by identifying $r_{0}$ with $r_{N}$,
and the factor $K(r)$ is even, non-negative and satisfies \cite{Kac-1963}
\begin{equation}
    \sum_{i=1}^{N} K_{N}(r_{i}) = 1.
\end{equation}

Taking the limit $N\to\infty$ so that
\begin{equation}
    K(r) = \lim_{N\to\infty} N K_{N}(Nr)
\end{equation}
and
\begin{equation}
    \label{eq:Kint}
    \int_{-1/2}^{1/2} K(r) dr = 1 ,
\end{equation}
we get the effective one-particle Hamiltonian
\begin{equation}
    H_{h}[f] = \dfrac{p^{2}}{2} + V_{r}[f](q,t) - h_{r}(t)\cos q,
\end{equation}
where
\begin{equation}
	\begin{split}
	    V_{r}[f](q,t)
	    = -& \int_{-1/2}^{1/2} dr' K(r-r') \\
	    &\times \iint \cos (q-q') f(q',p',r',t) dqdp.
	\end{split}
\end{equation}
The  single body distribution $f(q,p,r,t)$ evolves as 
the Vlasov equation \cite{alphaHMF-Vlasov} 
\begin{equation}
	\frac{\partial f}{\partial t} + \left\{H_{h}[f], f \right\} = 0.
\end{equation}

We consider the linear response for the uniform stable stationary 
configuration $f_{0}(q,p)$, which does not depend on the lattice point $r$. 
The external field modifies the state from $f_{0}(q,p)$
to $f_{0}(q,p)+f_{1}(q,p,r,t)$.
We define the modification of magnetization depending on $r$ as
\begin{equation}
    M_{r}^{1}(t) = \iint \cos q f_{1}(q,p,r,t) dqdp.
\end{equation}
Using the periodicity of the lattice,
we expand the modification $M_{r}^{1}(t)$,
the factor $K(r)$ and the external field $h_{r}(t)$ as
\begin{equation}
    M_{r}^{1}(t)
    = \sum_{n\in\mathbb{Z}} \tilde{M}_{n}^{1}(t) e^{2\pi inr},
    \quad
    K(r)
    = \sum_{n\in\mathbb{Z}} \tilde{K}_{n}e^{2\pi inr},
\end{equation}
and
\begin{equation}
    h_{r}(t)
    = \sum_{n\in\mathbb{Z}} \tilde{h}_{n}(t)e^{2\pi inr}.
\end{equation}
The Laplace transform with respect to time $t$ gives
\begin{equation}
    M_{r}^{1}(t)
    = \sum_{n\in\mathbb{Z}} \dfrac{e^{2\pi inr}}{2\pi}
    \int_{\Gamma} \dfrac{F(\omega)}{1-\tilde{K}_{n}F(\omega)}
    \hat{h}_{n}(\omega) e^{-i\omega t},
\end{equation}
where $\Gamma$ is the Bromwich contour,
$\hat{h}_{n}(\omega)$ is the Laplace transform of $\tilde{h}_{n}(t)$,
\begin{equation}
    F(\omega) = \int_{0}^{\infty} dt e^{i\omega t}
    \iint \cos q_{t} \{ \cos q, f_{0} \} dqdp,
\end{equation}
and $q_{t}$ is the solution to the canonical equation associated
with the Hamiltonian $H_{0}[f_{0}]$, which has zero external field.

Setting the external field as $h_{r}(t)\to h~(t\to\infty)$,
we have $\tilde{h}_{0}\to h$ and $\tilde{h}_{n}\to 0~(n\neq 0)$.
As discussed in \cite{LinearResponseII},
the surviving response is provided by
the pole of $\hat{h}_{0}(\omega)$ at $\omega=0$,
and other poles give dampings by the stability assumption of $f_{0}$.
The linear response is hence written by the same
stability functional $D^{\rm V}$ with the HMF model \cite{LinearResponseII} as
\begin{equation}
    M_{r}^{1}(t)
    \to \dfrac{F(0)}{1-F(0)}~h
    = \dfrac{1-D^{\rm V}}{D^{\rm V}}~h,
\end{equation}
where we used the fact $\tilde{K}_{0}=1$ from Eq.~\eqref{eq:Kint}.
From the above expression, we conclude
that the Vlasov susceptibility and the critical exponents
$\gamma_{\pm}^{\rm V}$ in the system \eqref{eq:HMFgeneral}
are the same with ones in the HMF model,
for uniform stable stationary configurations.

\section{Summary and discussions}
\label{sec:summary}

We investigated the critical exponent of susceptibility
in the Hamiltonian mean-field (HMF) model,
which is a mean-field ferro-magnetic model
and is approximately described by the Vlasov dynamics.
The classical mean-field theory gives the critical exponent $1$
both in the high- and low-energy phases,
but the linear response theory for the Vlasov systems
reveals that the exponent is the half of that of magnetization
in the low-energy phase, which is typically $1/4$.
This scaling is obtained not only in thermal equilibrium states,
but also in one-parameter families of quasistationary states
of the Jeans type, when the families have continuous phase transitions. 
Apart from the HMF model, the present theory can be applied 
to uniform stable stationary configurations of generalized systems,
whose interaction depends on distance between two lattice points
on which particles are.

Some remarks are discussed in the followings.

The first remark is about the validity of the linear response theory
close to critical points.
The theory assumes that $\delta f$ is vanishing when $h\to0$
with satisfying the condition $|h\chi^{\rm V}|\ll\ave{M}_{0}$.
The Vlasov susceptibility can be, therefore, computed by use of the Vlasov
linear theory even for a large $\chi^{\rm V}$
since it is computed in the limit of $h\to 0$.


The second 
remark is on the spectrum analysis used to compute susceptibilities
\cite{patelli-ruffo-12} in the inhomogeneous phase.
This methods does not consider whole the integrals of motions
but can be used to describe approximations of the linear theory for non-integrable systems.

The third 
remark is for the critical exponents in the
homogeneous phase. In the homogeneous equilibrium, 
the two susceptibilities satisfy 
$\chi^T = \chi^{\rm V}  = T_{\rm c}/(T- T_{\rm c})$ 
for $T > T_{\rm c}$.
Then, the isolated system shows the classical exponent,  
although the dynamics keeps an infinite number of Casimir invariants. 
Thus, Casimir constraints do not always bring about the {\it strange} critical exponent, 
and it depends on the initial equilibrium state.

Another remark is that the existence of invariants may break some thermodynamic laws. 
Indeed, local temperature in isolated crystalline clusters is not uniform by conservation of 
angular and translational momenta \cite{niiyama-07}.

We remark on other studies of the critical exponents
in the Vlasov framework.
Based on the theory on unstable manifolds of the Vlasov-Poisson equation \cite{crawford-95}, 
Ivanov et al. \cite{ivanov-05} found numerically that 
scaling laws are different from the ones predicted by the classical theory.
However, they start from unstable spatially homogeneous Maxwell distributions,
and no critical exponents are discussed in literature for stable states and QSSs.

We end this article remarking on observations in experiments.
Dynamical systems could get trapped in QSSs, therefore, measures on experimental setups 
will show the Vlasov prediction for systems with large enough number of particles.

\acknowledgements
The authors thank to S. Ruffo and C. Nardini
for useful discussions
and to M. M. Sano and M. Suzuki 
for informing about Ref.~\cite{suzuki-70}.
This work is inspired by discussions in the workshop held
in Centre Blaise Pascal, ENS-Lyon, August 2012.
SO acknowledges the support of 
Grants for Excellent Graduate Schools, 
the hospitality of University of Florence, 
and the JSPS Research Fellowships for Young Scientists 
(Grant No. 254728). 
YYY acknowledges the support of Grant-in-Aid for
Scientific Research (C) 23560069.


\begin{thebibliography}{99}

	 \bibitem{Kubo}
	R. Kubo,
    	J. Phys. Soc. Jpn. {\bf 12}, 570 (1957).    

    	\bibitem{chis1}
    	H. Falk,
    	Phys. Rev. {\bf 165}, 602 (1968).
    	
	\bibitem{chis2}
    	R. M. Wilcox,
    	Phys. Rev. {\bf 174}, 624 (1968).
	
   	\bibitem{mazur-69}
  	P. Mazur, Physica {\bf 43}, 533 (1969).
      
        \bibitem{suzuki-70}
    	M. Suzuki,
    	Physica {\bf 51}, 277 (1971).

	\bibitem{balescu}
        R. Balescu,
	Equilibrium and Nonequilibrium Statistical Mechanics, Wiley, New York, (1975).

	\bibitem{nicholson}
	D.R. Nicholson, 
	Introduction to Plasma Theory, John Wiley, (1983).

	\bibitem{LinearResponseI}
    	A. Patelli, 
    	S. Gupta, C. Nardini, and S. Ruffo,
    	Phys. Rev. E {\bf 85}, 021133 (2012).
      	
	\bibitem{LinearResponseII}
    	S. Ogawa and Y. Y. Yamaguchi,
    	Phys. Rev. E {\bf 85}, 061115 (2012).

	\bibitem{binney}
	J. Binney, S. Tremaine, 
	Galactic Dynamics, Princeton Series in Astrophysics, (1987).

	\bibitem{campa-dauxois-ruffo-09}
    	A. Campa, 
  	 T. Dauxois, and S. Ruffo,
  	Phys. Rep. {\bf 480}, 57 (2009).
	
        \bibitem{spohn-book}
      H. Spohn,
      Large Scale Dynamics of Interacting Particles,
      Springer-Verlag, Heidelberg, (1991).

	\bibitem{braun}
	W. Braun and K. Hepp, 
	Commun. Math. Phys. {\bf 56} 101 (1977).

	\bibitem{Jeans-1919}
	J. H. Jeans, 
	Mon. Not. R. Astron. Soc. {\bf 76}, 71 (1915).

	

       \bibitem{physicaa}
    	Y. Y. Yamaguchi, 
    	J. Barr\'e, F. Bouchet, T. Dauxois, and S. Ruffo,
    	Physica A {\bf 337}, 36 (2004).
    	

	\bibitem{spohn}
	J. Messer and H. Spohn, 
	J. Stat. Phys. {\bf 29} 561 (1982).

	\bibitem{inagaki-konishi-93}
    	S. Inagaki and T. Konishi, 
    	Publ. Astron. Soc. Jpn. {\bf 45}, 733 (1993).
    	
	\bibitem{antoni-ruffo-95}
    	M. Antoni and S. Ruffo,
    	Phys. Rev. E {\bf 52}, 2361 (1995).
	

	\bibitem{physicaaII}
    	J. Barr\'e,
    	F. Bouchet, T. Dauxois, S. Ruffo, and Y. Y. Yamaguchi,
    	Physica A {\bf 365}, 177 (2006).


        \bibitem{antoniazzi-07}
        A. Antoniazzi, D. Fanelli, S. Ruffo, and Y. Y. Yamaguchi,
        Phys. Rev. Lett. {\bf 99}, 040601 (2007).


	\bibitem{BOY10}
    	J. Barr\'e, A. Olivetti, and Y. Y. Yamaguchi,
    	J. Stat. Mech. P08002 (2010).

	\bibitem{ogawa13}
      	S. Ogawa, 
        Phys. Rev. E {\bf 87}, 062107 (2013).

	\bibitem{chavanis-11}
    	P. H. Chavanis, 
    	Eur. Phys. J. B {\bf 80}, 275 (2011).

        



	\bibitem{debuyl-mukamel-ruffo-11}
    	P. de Buyl,
    	D. Mukamel, and S. Ruffo,
    	Phys. Rev. E {\bf 84}, 061151 (2011).


	\bibitem{yoshida-93}
	H. Yoshida,
	Celest. Mech. Dyn. Astron. {\bf 56}, 27 (1993).

	\bibitem{Kac-1963} 
	M. Kac, G. E. Uhlenbeck, and P. C. Hemmer, 
	J. Math. Phys. {\bf 4}, 216 (1963).

	\bibitem{alphaHMF-Vlasov}
	R. Bachelard, T. Dauxois, G. De Ninno, S. Ruffo, and F. Staniscia,
	Phys. Rev. E {\bf 83}, 061132 (2011).
		
        
	\bibitem{patelli-ruffo-12}
    	A. Patelli and S. Ruffo, in preparation. 

	\bibitem{niiyama-07}
    	T. Niiyama, 
    	Y. Shimizu, T. R. Kobayashi, T. Okushima, and K. S. Ikeda,
    	Phys. Rev. Lett. {\bf 99}, 014102 (2007).

	\bibitem{crawford-95}
		J. D. Crawford,
		Phys. Plasmas {\bf 2}, 97 (1995).

	\bibitem{ivanov-05}
		A. V. Ivanov, S. V. Vladimirov, and P. A. Robinson, 
		Phys. Rev. E {\bf 71}, 056406 (2005).

\end{thebibliography}
\end{document}